# Water Eminence Scrutinizing Scheme Based On Zigbee and Wireless Antenna Expertise - A Study

[1]V.Karthikeyan, [2]Geethanjali.S, [3]Mekala.M, [4]Deepika.T

*Assistant Professor, Department of ECE, SVS College of Engineering, Coimbatore,*
*[2, 3, 4] Final Year Students, Department of ECE, SVS College of Engineering,*

[1]Karthick77keyan@gmail.com
[2]anjali.kanappan@gmail.com,
[3]mekala.mece@gmail.com
[4]breezedeepika@gmail.com

*Abstract*— **Wireless Sensor Network (WSN) is the essential structure of a water eminence monitoring by means of wireless sensor network (WSN) technology. To scrutinize water quality greater than different sites as a synchronized application, an estimable system structural design constituted by spread sensor nodes and a base station is suggested. The nodes and base stations are linked using WSN technology like Zigbee. Base stations are related via Ethernet. Design and execution of a prototype using WSN technology are the exigent work. Data's are identified by means of dissimilar sensors at the node plane to compute the parameters like pH, turbidity and oxygen quantity is transmitted via WSN to the support station. Information unruffled from the distant location is capable of displayed in diagram set-up as well as it is able to be calculated using dissimilar replication tools at the supporting station. The recent methods have benefits such as null amount carbon emission, low power utilization, more stretchy to put together at distant locations.**

## I. INTRODUCTION

The current technology is so maturity and we say as a century of globalization and the lot much else, but the same coin has second side too, that is nothing but the 21st century is said to be the century of inventions, warming, lack of self-confidence and helpless health factors! The critical and indispensable hurdle is increasing population do not comprise protected and clean drinking water. The condition is even worse in some emerging countries, where dirty or unhygienic water is being used for consumption without any proper & prior management. One of the reasons for this occurrence is the ignorance of public & management and the lack of water quality monitoring system and which creates severe health issues. [5] [6] This work started after taking into consideration the critical situation of the impure natural water resources in Malaysia. Keeping the water property so that it is always within a standard resolution for domestic usage is a fundamental assignment. As the country is constructing its evolution through industrialization, current water resources are flat to an exposure to pollution particularly from the industrial actions. It is a challenge in the enforcement feature as it is impractical for the powers that be to incessantly monitor the locality of water resources due to restriction in particular in manpower, services and the cost of tools. This often leads to a too late to be handled conditionally.

For that, it is important to have such a monitoring system with the uniqueness of self-sufficient, lower cost, consistent and flexible. The utilization of the processor in monitor the computation will make to reduce the man power in monitor the location and also reducing the cost.

The theme of the paper looks over on the utilization of a numerous sensors as a device to ensure the stage of water worth as a stand-in route of monitoring the state of affairs of the water resources. A number of sensors that are able to endlessly read a number of parameters that designate the water quality level such as chemical substances; conductivity, dissolved oxygen, pH, turbidity etc. will be able to monitor the entire quality level. As the monitoring is projected work to be carried out in a remote area with some degree of access, signal or information from the sensor unit will subsequently transmit wirelessly to the base monitoring station. There are more than a few nodes and a supporting station. Each node contains a collection of sensors and the nodes are extending in different water bodies. Information composed of sensors is transmitted to the base station via WSN channel. The base station is more often than not a PC with Graphic User Interface (GUI) for users to examine water quality data or alarm repeatedly when water quality detected when it is at below predetermined standards. The recorded data can be analyzed using various replication tools for future communication and actions. [1][3][7].

## II. VARIOUS METHODOLOGIES

1. Water Quality Monitoring System Using Zigbee Based Wireless Sensor Network

The constraints mixed up in the water worth resolve such as the pH level, turbidity and temperature are planned in the real time through the sensors that transmit the information to the reporting place or control room.

A. *Sensor Unit*

A sensor unit is essentially consists of a number of sensors used to notice the predestined parameters that identify the quality of water. In the projected work there are different types of sensors like pH sensor so as to senses the sourness or basicity of the water, temperature sensor and turbidity sensor based are based on the phototransistor to sense the temperature

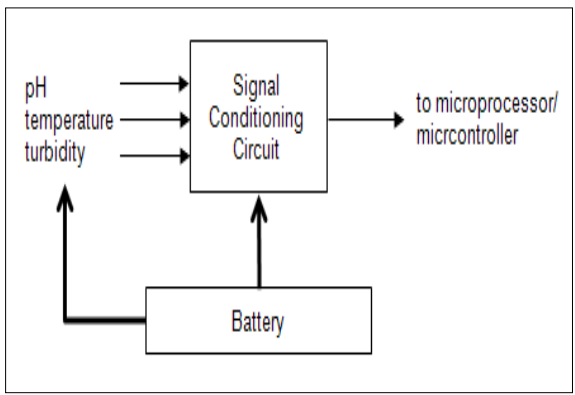

Fig 1. Block diagram of sensor unit

### B. Wireless Sensor Node

WSN is consisting of the sensor unit as mentioned. In a processor with a work of signal digitizing, information broadcast, networking management etc; and data line frequency transceiver for interactions at the physical layer. All of them contribute to a single battery as a power resource. The Fig. 2 illustrates the block diagram of the WSN NETWORK

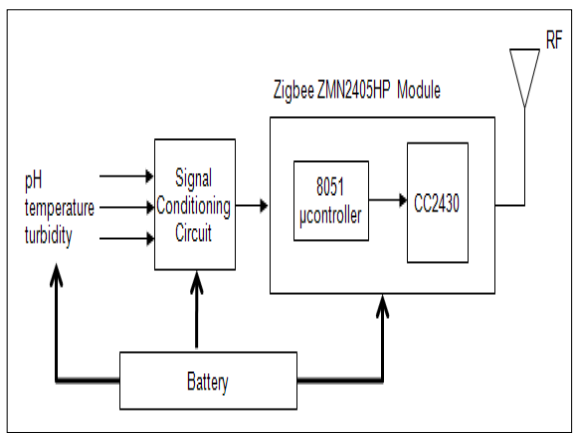

Fig 2. Zigbee based wireless sensor node

The important part is PIC microcontroller block which is re programmable and it operates as a conclusion device. As an end device sensor node, it is capable of only communicate with the router or coordinator to surpass the information from the sensor. An end device can only converse in some technique with the other end device in the course of the router or coordinator. The sensor node distinct as a router is in charge of routing information from other routers or end device to the host station or two supplementary routers closer to the microcontroller. The router can also be alive a data input device like the end device but in the case of real time purposes it is regularly used to extend the coverage distance of the monitoring system. There can be only one host station for the monitoring system. The host manager is answerable for the state of affairs through channel for the set of connections to use, the communication arrangement address of router and end device and maintaining the routing tables for the arrangement that are more prominent to transmit the information from one end device to another device and are in the same Zigbee system. 9V supply battery is provided and straightforwardly coupled to a 5V regulator before given to the Zigbee module.

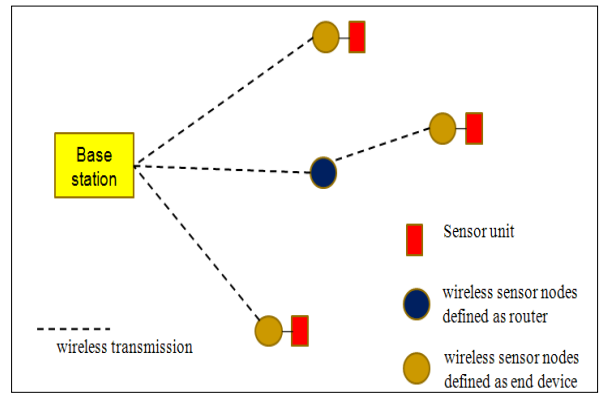

Fig.3. Zigbee based WSN monitoring system

### C. Base Monitoring Station

The data base station consists of a similar Zigbee module designed to receive the information's sent from the sensor nodes continuously. As the coordinator needs to all the time receiving data from the end devices, it is more often than not powered. The information received from the distant node sensors are again transmitted to the processor using the RS 232 interfacing probe and the received data's are displayed using the GUI on the base station.

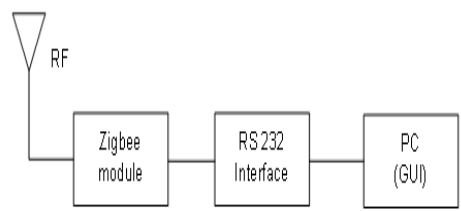

Fig. 4. Block diagram of components in base station

## III. REMOTE WATER POLLUTION MONITORING SYSTEM USING GSM

The sensor pH and turbidity will be set aside in the river water exterior and the information captured by the sensor will be specified in the PIC Micro controller, and after that the data's are transmitted wirelessly by means of Zigbee module. This technique consists of data base system and different sensor nodes. These types of sensors are powered and driven by Wind Piezoelectric power module; whereas the data composed between the node and the data station is designed using WSN technology.

The modules incorporated in the scheme architecture are as follows,

- PIC microcontroller
- ZIGBEE transceiver
- GSM

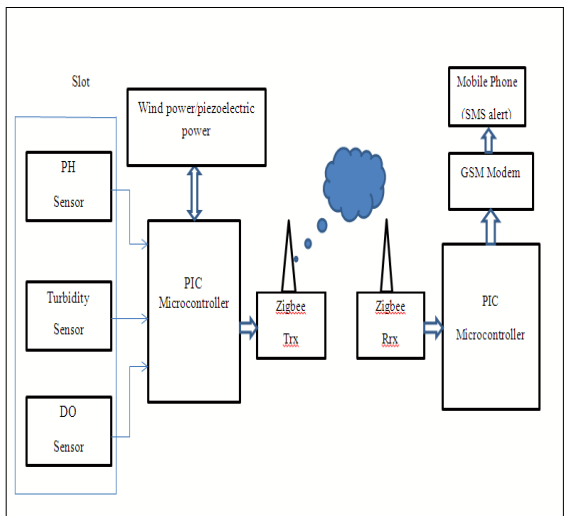

Fig. 5. Block diagram for transceiver

### A. PIC Microcontroller

A PIC microcontroller is a mainframe with inbuilt memory and RAM and you can utilize it directly to your projects. Consequently it saves you to construct a circuit that has detached external RAM, ROM and secondary chips.

### B. Zigbee Network

Zigbee is a design for a group of high level communication protocols using small, low power digital radios based on an IEEE 802 standard for personal area networks [4]. Applications comprised of wireless light switches, electrical meters with in-home-displays, and additional customer and industrialized tools that have need of low level-range wireless transmit of data at sensibly at low rates. The expertise defined by the Zigbee requirement is projected to be simpler and a smaller amount than other WPANs, such as Bluetooth. Zigbee has a distinct rate of 250 kbps best suited for periodic or irregular data or a single signal broadcast from a sensor or input device [4] [9].

### C. GSM

A GSM modem is a dedicated type of modem which accepts a SIM card, and operates more than a part to a mobile operator, presently similar to a mobile phone. As of the mobile operator perception, a GSM modem looks presently similar to a mobile phone.

## IV. WATER QUALITY WIRELESS SENSOR NETWORK (WQWSN)

The fundamental processes of this type of techniques are paying attention in studying analyzing the behavior of various parameters like Phosphorus, oxygen and temperature. The analyzed information's are put together at regular intervals of time (say for each an hour), and the statistics are collected and has to be automated as a retrievable data's at the end of the examined stage. The working of the WQWSN is to automate the quality of drinking water by monitoring the level of the standard level of the parameter. The theme of the method consists of the following steps, sampling the water quality at regular interval of time, transmitting the sampled data to the base station at every sampling time, resting the function for the limited time and wake up to repeat the procedure from the first step. These four steps might be giving as three-layered structural design depicted by Figure 6.

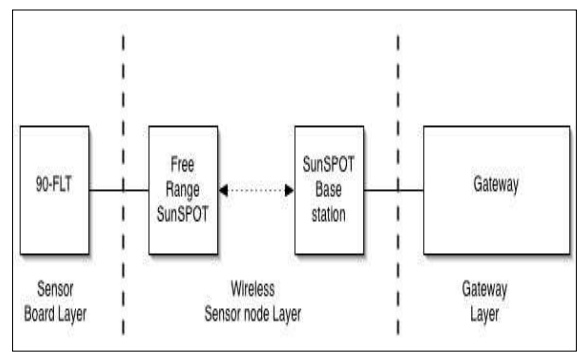

Fig 6. WQWSN system architecture

### A. Wireless sensor gateway layer

The gateway is one of the most significant device upon which the competence of the sensing action of a WSN depends. It collects all the data's received from the nodes in a database and makes this information obtainable usually via a wireless network. The conditions need a device intended approximately the subsequent constraints: low-power consumption, high storage capabilities, and pliable connectivity, low cost.

### B. Wireless sensor node layer

The layer of the wireless nodes was assembled from popular SunSPOT nodes [8] by means of a additional board. These types of nodes are used in Java TM technology to High level programming. The Sun SPOT drops into a power reduction mode ("shallow sleep") to decrease power utilization and lengthen battery life when all threads turn out to be inactive. The Sunspot nodes are of two types one used in the base station and other is used in the remote site. The supporting station unit communicates wirelessly with the free-range units and streams the data via a USB link. The SunSPOTs were chosen for their easiness of make use of and easy to get to interfaces.

### C. Water sensor board layer

All data's can be saved and downloaded directly to a processor via the standard RS232 port. The 90-FLT water eminence sensor is self-possessed of a major unit and some sensor probes: turbidity, pH and dissolved oxygen. The probes are the parts to be wrapped up in water, while the main unit has to be exterior. The main part in the method is the wireless sensor node. At the time of measurements the wireless nodes are awaken and fetch the most recent data's and send to the base station. Fetching data is the progression of communicating with the 90-FLT sensor. It includes waking up the sensor by transferring a particular sequence via the serial connection, reading data from the sensor, and rotating the sensor off by sensing one more string. Passing data to the SunSPOT base station, on the other hand, include assembling data in an additional reasonable format and sending it via radio to the host station. The SunSPOT base station is connected via USB to the entrance and data coming from the free-range SunSPOTs reach the SunSPOT base station via an 805.14 wireless link. The Java code in a row on the base station connects to the MySQL database that runs on the gateway and supplies the expected data in the database. The SunSPOT base station is also accountable for 1) sending the prototype files to the free-range SunSPOTs (containing the measurement times) and 2) synchronizing the clocks.

### V. PERFORMANCE EVALUATION:

We considered dissimilar presentation parameters. These include: (1) the power utilization and (2) the lifetime of the sensor system in unlike sensing scenarios. Whereas the power consumption is a limitation that measures the power utility of a wireless sensor mechanism, a wireless sensor network lifetime is articulated by the battery levels of its components. A lower power overwhelming wireless sensor network is preferable to a very much power overwhelming wireless sensor network.

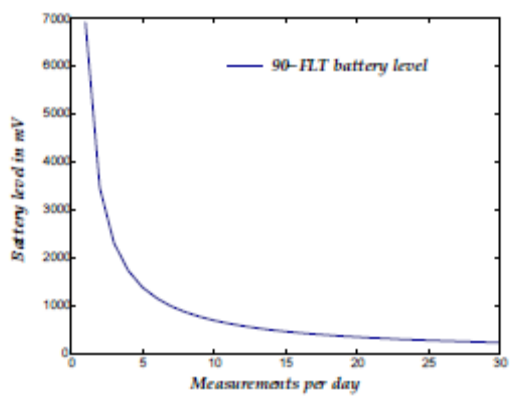

Fig.7. 90FLT lifetime

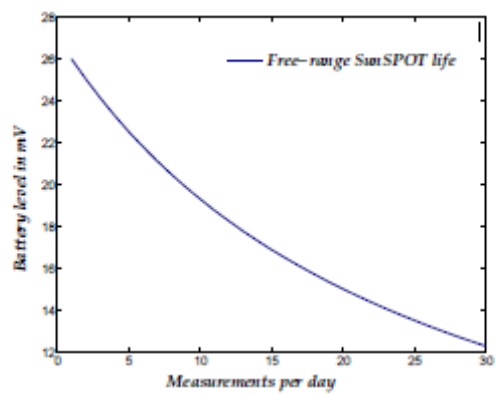

Fig 8. Sunspot lifetime

### VI. CONCLUSION

In the projected work, a narrative water eminence monitoring organization Zigbee based on wireless sensor network contributing small power utilization with high reliability is presented. They make use of large power WSN is appropriate for activities in industries concerning large region monitoring such as developing, constructing, taking out etc. An additional significant detail of this structure is the simple installation of the method where the base station can be located at the local habitation close to the object area and the monitoring task can be completed by any individual with minimal training at the opening of the system installation. The method is constituted by a base station and more than a few sensor nodes. The sensor nodes are powered by Piezoelectric/Wind Energy component, while data relation among the node and base station is realized by means of WSN technology (IEEE 802.15.4). On the node surface, water eminence data are collected by dissimilar sensors such as pH, DO and turbidity.

### REFERENCES

[1] Steven Silva, Hoang N Ghia Nguyen, Valentina Tiporlini and Kamal Alameh "Web Based Water Quality Monitoring with Sensor Network: Employing Zigbee and WiMax Technologies" by, 978-1-4577-1169-5/11/$26.00 ©2011 *IEEE*

[2] Cirronet, ZMN2405/*HP ZigbeeTM Module Developer's Kit User Manual, Rev A* 2007.

[3] F.Akyildiz lan, Su Weilian, Sankarasubramaniam Yogesh etc. A Survey on Sensor Networks 0163-6804/02 2002 IEEE.

[4] Li pengfei, Li jiakun,jing junfeng" Wireless Temperature Monitoring System Based on the Zigbee Technology"2nd *International Conference on computer Engineering and Technology* 2010.100

[5] F. Akyildiz, W. Su, Y. Sankarasubramaniam and E. Cayirci, "Wireless sensor networks: a survey," *Computer Networks*, Volume 38, Issue 4, pp 393-422, 2002.

[6] Tuan Le Dinh; Wen Hu; Sikka, P.; Corke, P.; Overs, L.; Brosnan, S,"Design and Deployment of a Remote Robust Sensor Network: Experiences from an Outdoor Water Quality Monitoring Network," *Local Computer Networks*, 32nd *IEEE Conference* on, pp 799-806, 2007.

[7] Puccinelli, D.; Haenggi, M., "Wireless sensor networks: applications and challenges of ubiquitous sensing," *Circuits and Systems Magazine, IEEE*, Vol.5, Issue 3, 2005.

[8] Sunspot world. Available: http://www.sunspotworld.com

[9] Deng yu, wang juan "Application of Remote Sensing Monitoring System in the Yellow River" 2nd *international conference on signal Processing system* 2010

[10] Rasin, Zulhani; Abdullah, Mohd Rizal "Water Quality Monitoring System Using Zigbee Based Wireless Sensor Network" *International Journal of Engineering & Technology*; Vol. 9 Issue 10, p24, 2009

[11] R. Henderson, A. Baker, K. Murphy, A. Hambly, R. Stuetz and S. Khan, "Fluorescence as a potential monitoring tool for recycled water systems: A review, "*Water Res.,* Vol. 43, pp. 863-881, 2009.

[12] K. Romer and F. Mattern, "The design space of wireless sensor networks, "*Wireless Communications, IEEE,* vol. 11, pp. 54-61, 2004.

[13] Seders L.A., Shea C.A., Lemmon M.D., Maurice P.A., Talley J.W. Lake Net: An Integrated Sensor Network for Environmental Sensing in Lakes. *Environ. Eng. SCI.*; 24:183–191 2007

[14] O'Flynn B., Martínez-Català F., Harte S., O'Mathuna C., Cleary J., Slater, C., Regan F., Diamond D., Murphy H. Smart Coast: A Wireless Sensor Network for Water Quality Monitoring. *32nd IEEE Conference on Local Computer Networks*,; Dublin, Ireland October 15–18,; pp. 815–816, 2007

[15] Jiang P. Survey on Key Technology of WSN-Based Wetland Water Quality Remote Real-Time Monitoring System *China Journal of Sensor & Actuators* 2007; 20:183–186

[16] Li Zhenan, Wang Kai, Liu Bo "Sensor-Network based Intelligent Water Quality Monitoring and Control" *International Journal of Advanced Research in Computer Engineering &Technology (IJARCET)* Volume 2, Issue 4, 2013

[17] Zulhani Rasin and 2Mohd Rizal Abdullah "Water Quality Monitoring System Using Zigbee Based Wireless Sensor Network" *International Journal of Engineering & Technology IJET-IJENS* Vol: 09 No: 10 2009

[18] Jing Fan, Nan Chen, Chang Song Xiang, Jin Long Wang "Study on Networking Technology Based on ZigBee with Water PH Monitoring for Wireless Mesh Sensor Network" *Advanced Materials Research* Vol: 630, pp 302-307, 2012